\documentclass{llncs}
\usepackage{graphicx}
\usepackage{todonotes}
\usepackage{comment}
\usepackage{enumitem}
\usepackage{microtype}
\usepackage{setspace}
\usepackage[hyphens]{url}
\usepackage{hyperref}
\title{Smart speaker design and implementation with biometric authentication and advanced voice interaction capability}
\author{Bharath Sudharsan, Peter Corcoran, Muhammad Intizar Ali}
\institute{Data Science Institute, National University of Ireland Galway. \\
\email{\{b.sudharsan1,peter.corcoran,ali.intizar\}@nuigalway.ie}}

\begin{document}
\maketitle
\begin{abstract}

Advancements in semiconductor technology have reduced dimensions and cost while improving the performance and capacity of chipsets. In addition, advancement in the AI frameworks and libraries brings possibilities to accommodate more AI at the resource-constrained edge of consumer IoT devices. Sensors are nowadays an integral part of our environment which provide continuous data streams to build intelligent applications. An example could be a smart home scenario with multiple interconnected devices. In such smart environments, for convenience and quick access to web-based service and personal information such as calendars, notes, emails, reminders, banking, etc, users link third-party skills or skills from the Amazon store to their smart speakers. Also, in current smart home scenarios, several smart home products such as smart security cameras, video doorbells, smart plugs, smart carbon monoxide monitors, and smart door locks, etc. are interlinked to a modern smart speaker via means of custom skill addition. Since smart speakers are linked to such services and devices via the smart speaker user's account. They can be used by anyone with physical access to the smart speaker via voice commands. If done so, the data privacy, home security and other aspects of the user get compromised. Recently launched, Tensor Cam's AI Camera, Toshiba's Symbio, Facebook's Portal are camera-enabled smart speakers with AI functionalities. Although they are camera-enabled, yet they do not have an authentication scheme in addition to calling out the wake-word. This paper provides an overview of cybersecurity risks faced by smart speaker users due to lack of authentication scheme and discusses the development of a state-of-the-art camera-enabled, microphone array-based modern Alexa smart speaker prototype to address these risks.

\end{abstract}

\section{Introduction} \label{Intro}
The recent advancements in technology (particularly IoT and AI) are having a great impact on our day to day life \cite{sudharsan2020avoid} \cite{sudharsan2022ris} \cite{sudharsan2020adaptive}. In a smart home scenario, multiple smart devices are interlinked and work in collaboration with each other to serve a common goal \cite{sudharsan2020edge2train} \cite{sudharsan2021toward} \cite{sudharsan2021globe2train}. Smart speakers are one amongst such smart devices that are being widely adopted by common users and becoming an integral part of smart homes. The AI assistants inbuilt within the recent smart speakers can understand voice-based commands and control complex integrated systems of a smart home. While voice-based commands provide an easy mechanism to interact with complex systems, they also introduce a security risk in terms of handing over control of systems to any user who has access to the smart speaker and can deliver voice-based commands. There is a strong need to introduce bio-metrics based authentication mechanisms for smart speakers to strengthen the security of integrated systems without compromising the rich user experience. Due to the lack of reliability in the existing voice authentications system, the ideal solution is to introduce additional authentication techniques.  

When a person claims to be the registered smart speaker user, there is a need to provide a factor to prove \emph{"the user is who she says she is"}. This factor can be providing the authentication system of the smart speaker with something the user knows (pin or password), or use something the user has (physical token) or something the user is (biometrics). Biometric authentication is best suited since authentication is a part of the user which makes the authentication process of smart speaker hands-free. Voice authentication analyzes the user's voice to verify identity based on the user's unique vocal attributes. Voice authentication is ideal for hands-free usage of standalone devices like smartphones, smart speakers and voice-based systems in an automobile since its integration is cost-effective, familiar and convenient for most users, less invasive (contactless) and more hygienic. But its downsides are, it is not as accurate as other biometric modalities \cite{awarebiometrics_2018_voice}, requires additional liveness detection system and background noise impacts the voice matching performance \cite{awarebiometrics_2018_voice}. Biometric authentication solutions such as Knomi \cite{awarebiometrics_2017_knomi} provide a family of biometric matching and liveliness detection algorithms that use both face \& voice for authentication. Likewise Sensory’s TrulyHandsfree \cite{a2014_trulysecure} uses proprietary face, voice recognition, and biometric fusion algorithms leveraging computer vision, speech processing and machine learning algorithms to provides on-device, almost instantaneous authentication. SDK's of such multi-modal authentication systems are suited to build applications for smartphones and tablets and not for smart speakers because of its low hardware specifications \cite{sudharsan2021ml} \cite{sudharsan2021porting} \cite{sudharsan2021sram}. 

Recently launched,  Tensor  Cam’s  AI  Camera,  Toshiba’s  Symbio,  Facebook’s Portal are camera-enabled smart speakers with  AI  functionalities \cite{sudharsan2020rce} \cite{sudharsan2021tinyml} \cite{sudharsan2021enabling}. Although they are camera-enabled, yet they do not have an authentication scheme in addition to calling out the wake-word. The modern Alexa smart speaker discussed in this paper is constructed using off the shelf hardware components (Raspberry Pi, ReSpeaker v2, Raspberry Pi camera, regular speaker). A biometrics-based authentication system for such Alexa smart speakers is designed by adding a camera module and introducing face recognition algorithms. This face recognition algorithm was trained using Deep Neural Network which can detect and identify human face for authentication. Additionally, it was able to identify and recognize faces during the human gaze, thus waking up Alexa only when a known face is recognized. To provide a seamless, full-duplex user-Alexa interaction, a microphone array with an on-board chip hosting DSP based speech algorithms was selected and used to capture, process and provide a noise suppressed voice feed to Alexa. Our proof of concept prototype demonstrates a rich user experience to interact with smart speakers by providing an extra layer of authentication and also facilitating improved voice interaction with the device.  

\section{Cybersecurity risks due to lack of authentication schemes in smart speakers}

Users start interacting with a regular Alexa smart speaker by waking up the Alexa AI voice assistant by calling out the “Alexa” wake-word, followed by regular dialogues based interaction. In the current scenario, a few Alexa devices support voice profiles \cite{a2019_amazoncom} to provide a personalized interaction experience with its supported features. For this, the user has to train Alexa using voice, followed by linking the trained voice with a corresponding Alexa user account.  But, this feature is only a voice-based user identification rather than authentication. Firstly, this existing voice-biometric feature is limited to a few Alexa supported features and does not act as a voice biometric authentication method for the whole smart speaker system. Secondly, it is proven that a similar voice might be able to fool Amazon and Google's voice recognition \cite{gebhart_2017_fooling} and also Google warns saying the fact that \emph{similar voice might be able to access your info} while the user is setting up voice recognition for the first time. According to a guide to the security of voice-activated smart speakers, for example, an ISTR Special Report published in 2017 \cite{a2017_a} and other similar research articles, the following are a few cybersecurity risks that the smart speaker user can get exposed to in the absence of user authentication scheme.

\begin{enumerate} [label=\alph*.]

\item \textbf{The curious child attack:} There is always a risk that a child can make a purchase via voice commands from the smart speaker without the knowledge of the linked account owner 

\item \textbf{Mischievous neighbor's tale:} If a neighbor wants to cause mischief. She/he could send commands to the smart speaker in ultrasonic frequencies which cannot be heard by humans but can be detected by smart speakers. 

\item \textbf{\emph{“This parrot keeps trying to buy food by speaking to Alexa” \cite{charlton_2018_this}:}} A parrot managed to successfully add items such as strawberries, light bulb, and kettle, etc. to owner's online shopping cart. Such activities could be avoided by using a pin, but the parrot could potentially learn and repeat the pin too.

\item \textbf{Talking television troubles:} Simply watching television or listening to the radio can Wake-up and interact with the smart speaker. 

\item \textbf{Physical access:}Anyone proximate to the smart speaker can wake it up, interact and extract information from the actual user’s calendar, reminder’s and other linked applications

\item \textbf{Biometric-override attack \cite{feng_2017_continuous}:} An attacker can inject voice commands \cite{panjwani_crowdsourcing} by replaying the previously recorded clip of the victim’s voice, or by impersonating the victim’s voice.

\item \textbf{Malicious commands:} Someone can generate malicious commands, which can be heard as garbled sounds by human ears, while the smart speakers interpret them as commands. Such commands can be embedded in online videos or TV advertisements to attack devices \cite{alanwar_2017_echosafe}. As smart speakers are always listening, they are susceptible to such security attacks \cite{sudharsan2021edge2guard} by devices which can generate malicious voices. Audio from television news triggered Amazon Echo to place orders for dollhouse \cite{liptak_2017_amazons}
\end{enumerate}

To address these issues, one possible existing method is to provide voice biometrics-based authentication for crucial third party applications such as calendars, email, banking, etc which are linked to Alexa. This can be done by integrating a third-party voice biometric API such as ArmorVox \cite {a2019_welcome}. By doing so, the raw voice file captured by smart speakers should be exposed to the third-party API, which certainly could cause the emergence of privacy and data security challenges in the future. Again these approaches do not provide an authentication method for the whole smart speaker system and still leaves the system exposed to risks. Unfortunately, since most state-of-the-art smart speakers do not have authentication methods, they are mostly ineffective in alleviating the mentioned issues. The prototype developed and described in this paper interacts with Alexa API by providing noise suppressed audio feed captured from a Microphone array and in addition, it is capable of performing Biometrics (facial recognition) based system wakeup in addition to calling out the Alexa wake-word. The importance of Biometrics-based authentication for smart speakers was discussed in this session and the development of such biometrics enabled smart speaker prototype will be discussed in upcoming sessions.

\section{Related Work}

VAuth \cite{feng_2017_continuous} is proposed for continuous authentication of voice assistants to defend against the threats caused due to the open nature of the smart speaker's voice channel. VAuth is a separate embedded system that is adopted on wearable devices, such as eyeglasses, earphones/buds, and necklaces. This system senses the body-surface vibrations of the user and matches it with the speech signal received by the voice assistant’s microphone. Although VAuth achieved 97\% detection accuracy, it is not a feasible solution to charge, maintain and carry this separate embedded system attached to the body of the user just to authenticate a smart speaker. Daon’s IdentityX \cite{a2019_daon} is a multi-modal, vendor agnostic identity services platform that provides additional biometrics-based authentication using a smartphone only while using financial services apps via Alexa. This process involves the use of a secondary gadget (smartphone) and still not an authentication scheme for the entire smart speaker system which leaves Alexa exposed to risks discussed in Section \ref{Intro}. EchoSafe \cite{alanwar_2017_echosafe} is a sonar-based defense against the attacks which occur due to malicious voice commands from nearby devices during user unoccupied periods. Here, when the user sends a critical command to the smart speaker, an audio pulse is sent from the smart speaker followed by post-processing to determine if the user is present in the room. They have claimed EchoSafe system can detect the user’s presence during critical commands with 93.13\% accuracy. EchoSafe is a solution only for attacks via malicious voice commands and not suited for other vulnerabilities. 

\section{Overview of biometric authentication and speech algorithm based smart speaker} 

\begin{figure}[t]
\centering
 \includegraphics[width=12cm,height=7cm,keepaspectratio]{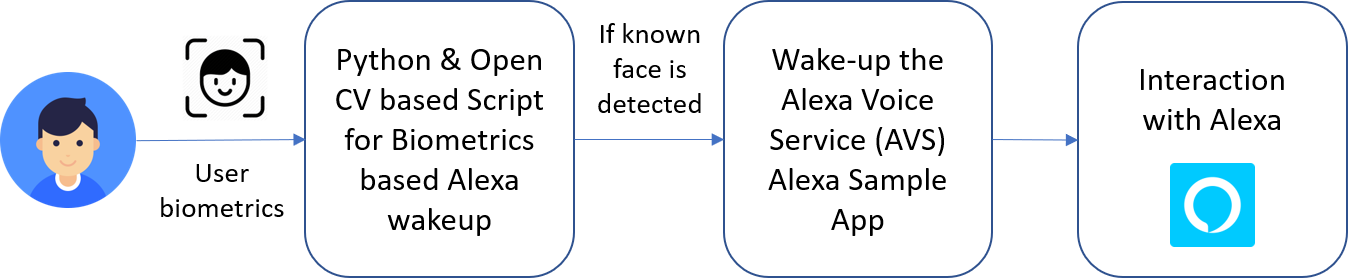}
\caption{High-level system diagram of Alexa smart speaker prototype with biometrics-based wakeup}
\label{fig:1}  
\end{figure}

\begin{figure}
\centering
 \includegraphics[width=10cm,height=8cm,keepaspectratio]{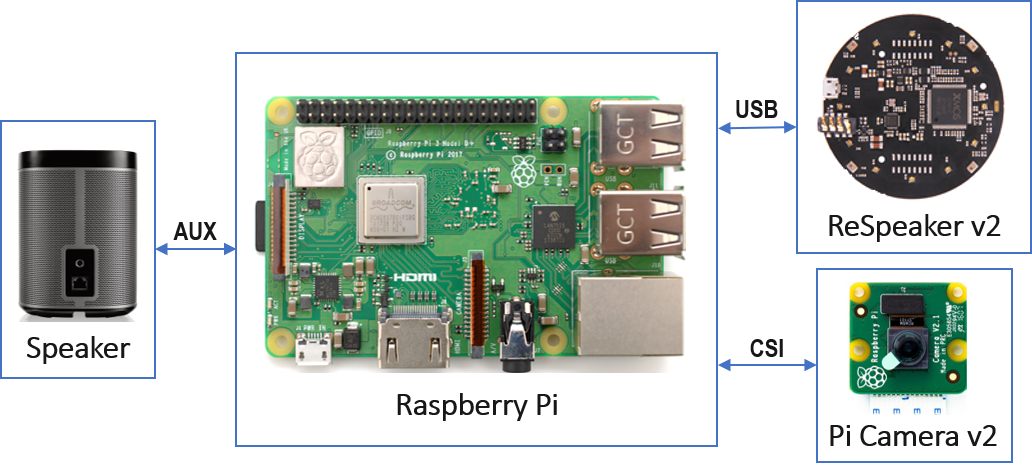} 
\caption{Hardware prototype of smart speaker system}
\label{fig:2}  
\end{figure}

The first objective of this work is to provide a face biometrics-based authentication scheme for the entire smart speaker system. To perform this, a camera module is added to the smart speaker prototype as shown in Fig. \ref{fig:2}. The lack of authentication schemes in regular smart speakers provides an open door to access user's private information by anyone in its vicinity. Since the prototype discussed in this paper has a camera module and also is equipped with face recognition based Alexa wakeup scripts, it provides an extra layer of authentication. As shown in Fig. \ref{fig:1}, the registered user has to first gaze at the camera to authenticate the system, then call out the Alexa wake word and start the regular dialogue-based interaction with Alexa. Section \ref{BAW} discusses the algorithms involved to wake up the system when known face gazes at the prototype. The second objective is to capture and provide high-quality noise suppressed voice input to Alexa for achieving a seamless, full-duplex user-Alexa speech interaction. To perform this, the ReSpeaker v2 microphone array is used here rather than a single microphone since it can segregate speech from noise. Also, this mic-array has an inbuilt high-performance processor loaded with on-chip advanced DSP (Digital Signal Processing) based speech algorithms which enables users to interact with Alexa up to five meters or further from the smart speaker, interact while walking around the room, etc. This mic array's role and benefits of using it to capture, process and provide voice input for Alexa are discussed in Section \ref{RB}. The third objective is to improve the user experience by making sure the smart speaker is not activated accidentally when wake-word is not called out and also make sure the Alexa wake word is spotted from the input audio streams with high accuracy. To perform this a third-party wake word engine as discussed in Section \ref{Sensory} is integrated with the Alexa Voice Service C++ SDK as discussed in Section \ref{C++}.

\subsection{Hardware components of the smart speaker prototype}
This modern smart-speaker prototype is constructed using commercial off the shelf advanced microphone array with inbuilt DSP (ReSpeaker v2), camera module (Raspberry Pi camera), and a regular speaker interfaced to a single board computer as shown in Fig. \ref{fig:1}.

\space
\begin{enumerate} [label=\alph*.]
\item\textbf{Selection of Single Board Computer: }BeagleBone Black, Orange Pi 3, LattePanda 2G/32G and Banana PI M4 are the SBC's of our interest. The Raspberry Pi 3 model b+ (Pi 4 not yet released) is chosen considering its form factor, price-performance balance, low-power consumption, compatibility with off the shelf devices, community-created guides, tutorials, and support. As illustrated in Fig. \ref{fig:1}, python scripts written leveraging external libraries are deployed on this Raspberry Pi Linux SBC. The scripts deployed here are responsible for waking up the Alexa Sample App when a known face is recognized from live frames captured

\item\textbf{Selection of camera unit: }For real-time computer vision applications, the Raspberry Pi Camera V2 is preferred since it is capable of 1080p 30fps video encoding and 5MP stills quality. Since the camera is connected directly to the GPU via CSI connector as shown in Fig.\ref{fig:1}, there is only a little impact on Pi’s CPU, leaving it available for other processing. Most cost-effective web cameras do not have a built-in encoding like the Pi camera. Hence, web cameras use additional CPU resources causing the reduced overall performance of the system

\item\textbf{Selection of microphone array: }Microphone is the crucial part of a smart speaker system. Since we require pre-processing of sound using speech algorithms, the focus is on a microphone array with built-in advanced DSP algorithms. ReSpeaker v2, Matrix Creator, PS3 eye, Conexant 4-mic development kit, MiniDSP UMA-8, Microsemi AcuEdge ZLK38AVS are the microphone arrays of our interest. ReSpeaker v2 has a good success rate for hot word detection when the distance is increased and tested in a silent room, a room with white noise and room with background music \cite{rouchon_2017_benchmarking}. The PS3 Eye has the edge over ReSpeaker v2, but ReSpeaker v2 is chosen for this project because the Raspberry Pi camera with CSI interface has better support for Open CV environment than the PS3 Eye camera. The second reason is, the ReSpeaker has a Pixel Ring of 12 RGB LED’s which can be used for visual feedback in addition to the speaker unit.

\end{enumerate}
\subsection{Speech algorithms based microphone array for advanced voice interaction capability } \label{RB}

\begin{figure}
    \centering
     \includegraphics[width=12cm,height=9cm,keepaspectratio]{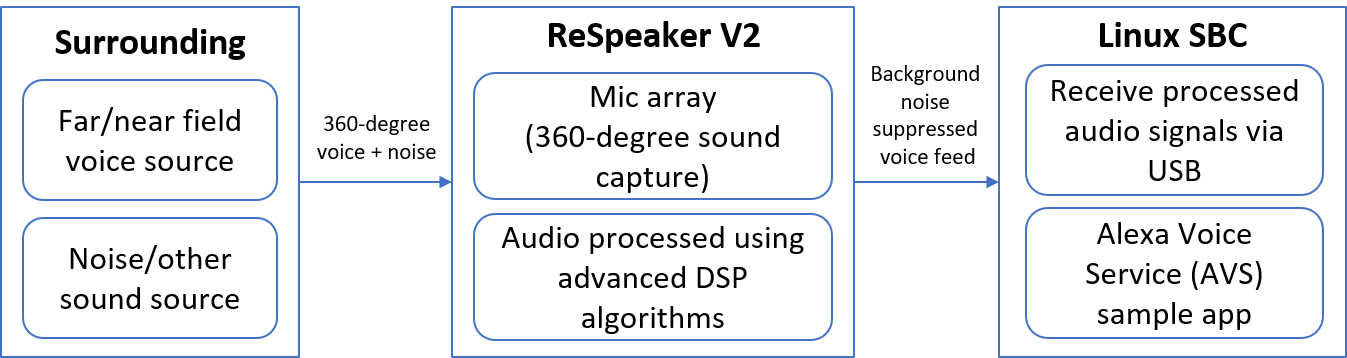}
    \caption{Flow diagram to illustrate microphone array's role: Capture and provide speech algorithm processed voice feed to Alexa Voice Service }
    \label{fig:3}
\end{figure}

The firmware on the XVF-3000 Chip (present on the ReSpeaker v2 hardware) produces six-channel mic outputs via USB to the Linux system. Channel zero contains audio which is processed using advanced DSP algorithms. Channels one to four contains raw data from the microphones corresponding to the channel number. Channel five provides raw audio which is a combination of all raw audio signals from four microphones on the ReSpeaker v2. A high-level illustration of this mic array’s role is shown in Fig.\ref{fig:3}. Here, the audio feed from channel zero is used for wake word spotting and also fed as voice input to the Alexa. The benefits of using ReSpeaker v2 with Alexa are listed below.

\begin{enumerate} [label=\alph*.]
\item \textbf{Far-field voice capture:} Wake-up and interact with the smart speaker by capturing and processing raw microphone inputs at distances of up to five meters or further.
\item \textbf{USB Audio Class 1.0 (UAC 1.0)}: USB audio is used to send digital music from the Raspberry Pi to the digital to analog converter (DAC) inbuilt on the ReSpeaker v2. Class 1.0 can send up to a maximum of 24- bit/96kHz hi-res files. By utilizing this we can bypass the internal sound-card of Raspberry Pi and allow the USB DAC to play audio response from Alexa with much better quality.
\item  \textbf{Twelve programmable RGB LED Pixel-ring:} The RGB LED pixel ring on the ReSpeaker v2 is utilized to visually point the direction of speech signal arrival (source). Pixel ring library is used to address the LED pixels via the USB interface to change color and brightness according to requirements from the main program.
\item \textbf{Digital Signal Processing algorithms on 
ReSpeaker V2:}
\begin{enumerate} [label=\roman*.]
\vspace{1 mm}
\item \textbf{Beamforming:} All MEMS microphones have an omnidirectional pickup response. It means, their response is the same for sound coming anywhere from around the microphone. Directional response or a beam pattern can be formed by configuring multiple microphones in an array. Thus, enabling us to detect and track the position of the voice of the smart speaker user across the room. As the smart speaker user interacts with the smart speaker and walks around the room, the angle of the microphone beam adjusts automatically to track their voice. Hence, it is effectively possible to point towards the user’s direction and suppress noise or reverberation signals from other directions.

\item \textbf{Noise suppression:} In acoustic beamforming, the spatial relationship of the microphones in the microphone array achieves active microphone noise suppression and control. If the direction of the sound source relative to the microphone array is known, then an acoustic beamformer can be designed to pass signals coming from the sound source of interest and filter out sound signals picked up from other different directions. This approach to microphone array noise reduction is most applicable to a situation in which one person’s voice needs to be heard when multiple people are talking. Noise suppression removes the stationary (point-noise) and non-stationary background sounds.

\item \textbf{De-reverberation:}  In any room, one’s voice will reverberate (reflect) off hard surfaces around the room, e.g. a window or TV screen. De-reverberation removes these reflections and cleans up the voice signal. 

\item \textbf{Acoustic Echo Cancellation:} While interacting with electronic devices, in some cases, users hear their voice (sometimes with a significant delay). This experience is known as an acoustic echo. Controlling and canceling acoustic echo is essential for voice-based systems such as smart speakers. For example, if the smart speaker user is watching a film on a TV with minimal volume and simultaneously gives voice input to the smart speaker, now the microphones will capture both the user’s voice and the sound of the film (the acoustic echo). This acoustic Echo is canceled from the voice input so that text from captured audio can be extracted with better accuracy.

\end{enumerate}

\subsection{Biometric authentication based Alexa wakeup} \label{BAW}
As illustrated in Fig. \ref{fig:5}. When the face recognition script is run, faces are detected from the live frames captured from Pi camera and a 128-d face embedding is computed via a deep metric network for the detected face. Then the computed 128-d face embedding is compared with a known database of already computed face encodings of registered faces to successfully recognise faces from the live frame. Once a known face is recognised, this script wakes up Alexa and simultaneously the ReSpeaker’s RGB LED Pixel-ring provides visual feedback to the user by turning green. Before running this face recognition script. A sub-script as shown in Fig. \ref{fig:4} has to be run in order to encode 128-d vectors for faces in the dataset (directory with .jpg files of faces) \& store the encodings in a .pickle file, which is later used as a database (while running main face recognition script) to compare detected faces from live frames and check for a match. Since Pi has limited computation power, memory \& GPU, its resource has to be left free for other scripts to run. Hence more powerful algorithms such as Eigenfaces and LBPs (Local Binary Patterns) which can achieve frame rates greater than 10 FPS was not used.

\begin{figure}
    \centering
   \includegraphics[width=10cm,height=9cm,keepaspectratio]{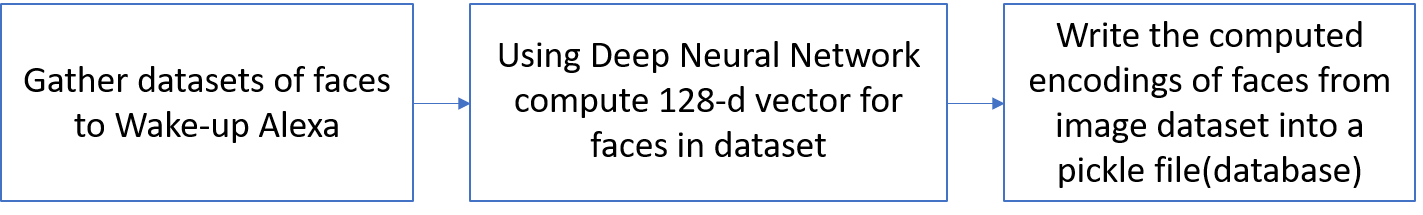}
    \caption{Flow diagram for preparing database of faces}
    \label{fig:4}
\end{figure}

\begin{figure}
    \centering
  \includegraphics[width=10cm,height=9cm,keepaspectratio]{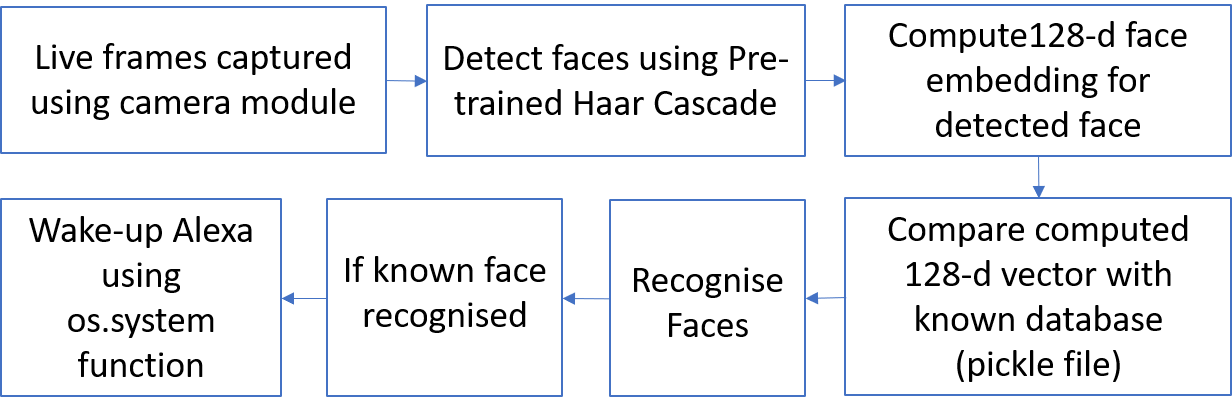}
    \caption{Flow diagram of face recognition script}
    \label{fig:5}
\end{figure}

\subsection{ Alexa Voice Service C++ SDK} \label{C++}
Parallel to setting up the smart speaker hardware and deploying the scripts, the Pi has to be registered as a device at Amazon developer console \& a security profile has to be created. Follow the detailed step-by-step instructions for Cloud side setup at \cite{alexa_2019_alexaavsdevicesdk} and provide the path of the generated config.json while building the Alexa AVS Sample App from its SDK. This C++ libraries based AVS device SDK enables us to integrate Alexa into the smart speaker prototype. The interaction of smart speaker with AVS is performed using this Alexa Sample App which is built for Raspberry Pi from the official SDK. Before proceeding with Alexa AVS C++ SDK, Python version of Alexa Voice Service app \cite{respeaker_2018_respeakeralexa} was tested with Raspberry Pi \& ReSpeaker v2. Following results were observed.
\begin{enumerate} [label=\alph*.]
    \item After interacting with Alexa for quite some time, Alexa's voice turned blurred \& muffled. It gets resolved only after restarting the Pi 
    \item After spotting the Alexa wake-word, there is a delay for a minimal duration (approx. 0.5 seconds) only after which audio is streamed to Alexa cloud   
\end{enumerate}

As mentioned in \cite{alexa_2019_alexaavsdevicesdk} and in Fig.\ref{fig:6} multiple components comprise the C++ AVS SDK through which the audio data flows. Initially, signal processing algorithms are applied to input and output audio channels to produce processed, clear audio. If the raw audio data from four microphones of ReSpeaker are provided as input then this third-party Audio Signal Processor combines and provides a single audio stream to the next component in the architecture. But here, we are already providing a single channel audio stream which is processed by the DSP on the XVF-3000 chip of ReSpeaker v2. The remaining subparts of the architecture perform its functionality as mentioned in \cite{alexa_2019_alexaavsdevicesdk} and Fig.\ref{fig:6}. Snowboy from KITT.ai and Sensory\cite{a2014_home} are two third-party wake word engines, either one of which has to be a part of the SDK build to spot the Alexa wake-word from the input streams to provide hands-free interaction. Both these engines were tested with this ReSpeaker v2 based Alexa smart speaker setup. Snowboy, wake work engine was selected and used as a plugin for building the Alexa AVS sample app since it only consumes less than 8 \% of Raspberry Pi's CPU and had a better success for wake word detection.

\begin{figure}[t]
\centering
 \includegraphics[width=10cm,height=11cm,keepaspectratio] {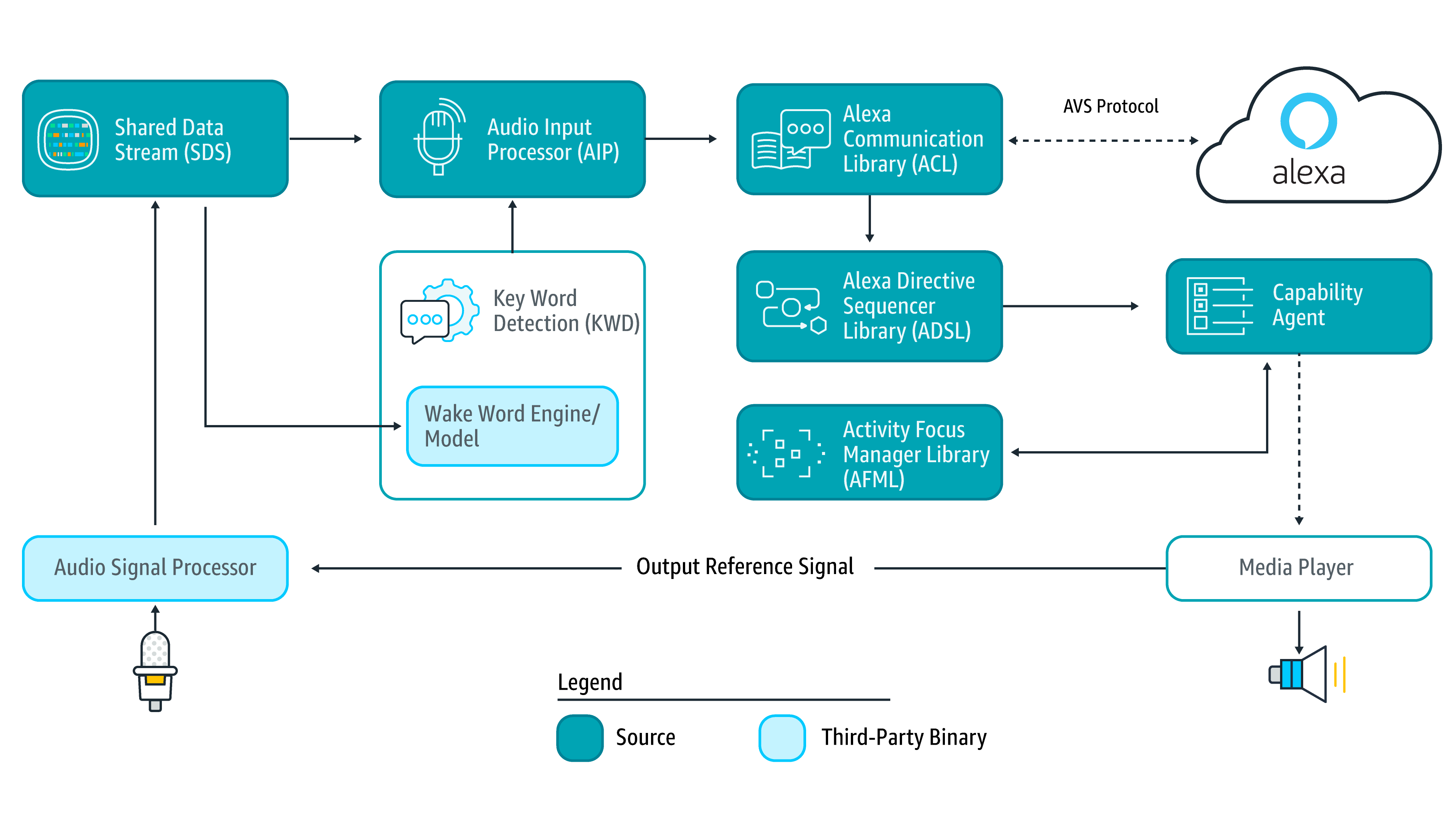}
  \caption{Data flow between components of AVS Device C++ SDK \cite{alexa_2019_alexaavsdevicesdk}}
\label{fig:6}       
\end{figure}

\subsection{ Snowboy wake-word engine to spot Alexa wake-word } \label{Sensory}
Snowboy engine ensures the smart speakers are not activated accidentally when wake-word is not called out. The accuracy of the wake word detection engines is measured by plotting false alarm per hour (a number of false positives) vs miss detection rates (percentage of wake word utterances an engine rejects incorrectly). The ROC curves of four different wake-word detection engines is shown in Fig.\ref{fig:7}. Here, Snowboy wake-word engine has the lowest miss detection rate and it is more accurate than other engines. Reasons for integrating wake-word engines with voice based AI assistants and smart speakers \cite{alirezakenarsari_2018_yet}.  

\begin{enumerate} [label=\alph*.]
    \item Privacy: Microphones does not have to listen always 
    \item Cost: Impractical \& expensive when data is streamed to cloud all the time 
    \item Power consumption: Voice assistants are run on smartphones, wearables \& smart speakers where maximum standby time is expected
\end{enumerate}

\begin{figure}
\centering
 \includegraphics[width=8cm,height=8cm,keepaspectratio] {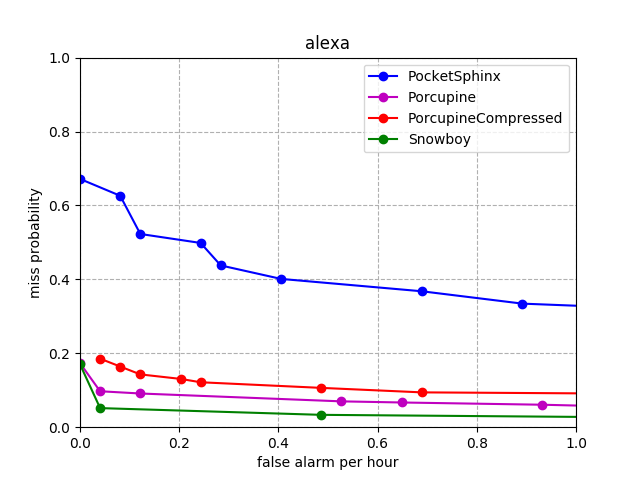}
  \caption{ROC curves for popular wake-word engines \cite{picovoice_2018_picovoicewakewordbenchmark}}
\label{fig:7}       
\end{figure}

 \end{enumerate}
 
\section{Conclusion and Discussion}

Over 50\% of Irish population are expected to own a regular smart speaker by 2023 \cite{taylor_2019_over} and its predicted that smart speaker ownership will overtake that of tablets globally by 2021 \cite{taylor_2019_over}. Likewise, soon camera-enabled smart speakers will replace regular smart speakers and become an integral part of our daily life. This paper provided an overview of cybersecurity risks faced by smart speaker users due to lack of authentication scheme and discussed the development of a state-of-the-art camera-enabled, microphone array based modern Alexa smart speaker prototype to address these risks. In addition to this biometrics based system wake up and microphone array-based interaction. Since this smart speaker prototype is a camera-enabled Linux-based system, it is capable to host custom skills which can perform audio processing and computer vision based tasks when requested by the user. We plan to extend our existing work for multiple use-cases requiring voice commands such as smart enterprises (online meetings) and other digital voice assistants \cite{sudharsan2019ai} \cite{sudharsan2019microphone} \cite{sudharsan2021owsnet}. 

\section*{Acknowledgements}
This publication has emanated from research supported by a research grant from Science Foundation Ireland (SFI) under Grant Number SFI/16/RC/3918 (Confirm), and SFI/12/RC/2289\_P2 (Insight) co-funded by the European Regional Development Fund.
\bibliography{aics-sample.bib}

\begin{thebibliography}{10}
\providecommand{\url}[1]{\texttt{#1}}
\providecommand{\urlprefix}{URL }

\bibitem{a2017_a}
A guide to the security of voice-activated smart speakers an istr special
  report,
  \url{https://www.symantec.com/content/dam/symantec/docs/security-center/white-papers/istr-security-voice-activated-smart-speakers-en.pdf}

\bibitem{picovoice_2018_picovoicewakewordbenchmark}
Picovoice benchmark, \url{https://github.com/Picovoice/wakeword-benchmark}

\bibitem{a2014_home}
Home | sensory (2014), \url{https://www.sensory.com/}

\bibitem{a2014_trulysecure}
Trulysecure | sensory (2014),
  \url{https://www.sensory.com/products/technologies/trulysecure/}

\bibitem{a2019_amazoncom}
Amazon.com help: About alexa voice profiles (2019),
  \url{https://www.amazon.com/gp/help/customer/display.html?nodeId=202199440}

\bibitem{a2019_daon}
Daon | multi-factor mobile biometric authentication | identityx platform - daon
  (07 2019), \url{https://www.daon.com/}

\bibitem{a2019_welcome}
Welcome - armorvox (2019), \url{https://cloud.armorvox.com/}

\bibitem{alanwar_2017_echosafe}
Alanwar, A., Balaji, B., Tian, Y., Yang, S., Srivastava, M.: Echosafe.
  Proceedings of the 1st ACM Workshop on the Internet of Safe Things -
  SafeThings'17  (2017)

\bibitem{alexa_2019_alexaavsdevicesdk}
{Alexa}: alexa/avs-device-sdk, \url{https://github.com/alexa/avs-device-sdk}

\bibitem{awarebiometrics_2017_knomi}
Biometrics, A.: Knomi™ - mobile biometric authentication framework - aware
  biometrics (2017),
  \url{https://www.aware.com/knomi-mobile-biometric-authentication/}

\bibitem{awarebiometrics_2018_voice}
Biometrics, A.: Voice authentication technology - aware biometrics software
  (2018), \url{https://www.aware.com/voice-authentication/}

\bibitem{charlton_2018_this}
Charlton, A.: This parrot keeps trying to buy food by speaking to alexa (12
  2018),
  \url{https://www.gearbrain.com/parrot-uses-amazon-alexa-2623633611.html}

\bibitem{feng_2017_continuous}
Feng, H., Fawaz, K., Shin, K.G.: Continuous authentication for voice
  assistants. Proceedings of the 23rd Annual International Conference on Mobile
  Computing and Networking - MobiCom '17  (2017),
  \url{https://dl.acm.org/citation.cfm?doid=3117811.3117823}

\bibitem{gebhart_2017_fooling}
Gebhart, A.: Fooling amazon and google's voice recognition isn't hard (11
  2017),
  \url{https://www.cnet.com/news/fooling-amazon-and-googles-voice-recognition-isnt-hard/}

\bibitem{alirezakenarsari_2018_yet}
Kenarsari, A.: Yet another wake-word detection engine (04 2018),
  \url{https://medium.com/@alirezakenarsarianhari/yet-another-wake-word-detection-engine-a2486d36d8d4}

\bibitem{liptak_2017_amazons}
Liptak, A.: Amazon’s alexa started ordering people dollhouses after hearing
  its name on tv (01 2017),
  \url{https://www.theverge.com/2017/1/7/14200210/amazon-alexa-tech-news-anchor-order-dollhouse}

\bibitem{panjwani_crowdsourcing}
Panjwani, S., Prakash, A.: Crowdsourcing attacks on biometric systems,
  \url{https://www.usenix.org/system/files/conference/soups2014/soups14-paper-panjwani.pdf}

\bibitem{respeaker_2018_respeakeralexa}
{Respeaker}: Alexa, \url{https://github.com/respeaker/Alexa}

\bibitem{rouchon_2017_benchmarking}
Rouchon, C.: Benchmarking microphone arrays: Respeaker, conexant, microsemi
  acuedge, matrix creator, minidsp… (08 2017),
  \url{https://medium.com/snips-ai/benchmarking-microphone-arrays-respeaker-conexant-microsemi-acuedge-matrix-creator-minidsp-950de8876fda}

\bibitem{sudharsan2020adaptive}
Sudharsan, B., Breslin, J.G., Ali, M.I.: Adaptive strategy to improve the
  quality of communication for iot edge devices. In: 2020 IEEE 6th World Forum
  on Internet of Things (WF-IoT). pp. 1--6. IEEE (2020)

\bibitem{sudharsan2020edge2train}
Sudharsan, B., Breslin, J.G., Ali, M.I.: Edge2train: A framework to train
  machine learning models (svms) on resource-constrained iot edge devices. In:
  Proceedings of the 10th International Conference on the Internet of Things.
  pp. 1--8 (2020)

\bibitem{sudharsan2020rce}
Sudharsan, B., Breslin, J.G., Ali, M.I.: Rce-nn: a five-stage pipeline to
  execute neural networks (cnns) on resource-constrained iot edge devices. In:
  Proceedings of the 10th International Conference on the Internet of Things.
  pp. 1--8 (2020)

\bibitem{sudharsan2021globe2train}
Sudharsan, B., Breslin, J.G., Ali, M.I.: Globe2train: A framework for
  distributed ml model training using iot devices across the globe. In: 2021
  IEEE SmartWorld, Ubiquitous Intelligence \& Computing, Advanced \& Trusted
  Computing, Scalable Computing \& Communications, Internet of People and Smart
  City Innovation (SmartWorld/SCALCOM/UIC/ATC/IOP/SCI). pp. 107--114. IEEE
  (2021)

\bibitem{sudharsan2021ml}
Sudharsan, B., Breslin, J.G., Ali, M.I.: Ml-mcu: A framework to train ml
  classifiers on mcu-based iot edge devices. IEEE Internet of Things Journal
  (2021)

\bibitem{sudharsan2019microphone}
Sudharsan, B., Chockalingam, M.: A microphone array and voice algorithm based
  smart hearing aid. arXiv preprint arXiv:1908.07324  (2019)

\bibitem{sudharsan2019ai}
Sudharsan, B., Kumar, S.P., Dhakshinamurthy, R.: Ai vision: Smart speaker
  design and implementation with object detection custom skill and advanced
  voice interaction capability. In: 2019 11th International Conference on
  Advanced Computing (ICoAC). pp. 97--102. IEEE (2019)

\bibitem{sudharsan2021owsnet}
Sudharsan, B., Malik, S., Corcoran, P., Patel, P., Breslin, J.G., Ali, M.I.:
  Owsnet: Towards real-time offensive words spotting network for consumer iot
  devices. In: 2021 IEEE 7th World Forum on Internet of Things (WF-IoT). pp.
  83--88. IEEE (2021)

\bibitem{sudharsan2021toward}
Sudharsan, B., Patel, P., Breslin, J., Ali, M.I., Mitra, K., Dustdar, S., Rana,
  O., Jayaraman, P.P., Ranjan, R.: Toward distributed, global, deep learning
  using iot devices. IEEE Internet Computing  25(03),  6--12 (2021)

\bibitem{sudharsan2021enabling}
Sudharsan, B., Patel, P., Breslin, J.G., Ali, M.I.: Enabling machine learning
  on the edge using sram conserving efficient neural networks execution
  approach. In: Joint European Conference on Machine Learning and Knowledge
  Discovery in Databases. pp. 20--35. Springer (2021)

\bibitem{sudharsan2021sram}
Sudharsan, B., Patel, P., Breslin, J.G., Ali, M.I.: Sram optimized porting and
  execution of machine learning classifiers on mcu-based iot devices: demo
  abstract. In: Proceedings of the ACM/IEEE 12th International Conference on
  Cyber-Physical Systems. pp. 223--224 (2021)

\bibitem{sudharsan2021porting}
Sudharsan, B., Patel, P., Wahid, A., Yahya, M., Breslin, J.G., Ali, M.I.:
  Porting and execution of anomalies detection models on embedded systems in
  iot: Demo abstract. In: Proceedings of the International Conference on
  Internet-of-Things Design and Implementation. pp. 265--266 (2021)

\bibitem{sudharsan2022ris}
Sudharsan, B., Rahul, P.S., Yadav, P., Gupta, S.K., Kumar, V., Nguyen, D.D.,
  Ali, M.I., Breslin, J.G.: Ris-iot: Towards resilient, interoperable, scalable
  iot. In: 2022 ACM/IEEE 13th International Conference on Cyber-Physical
  Systems (ICCPS). pp. 296--297. IEEE (2022)

\bibitem{sudharsan2021tinyml}
Sudharsan, B., Salerno, S., Nguyen, D.D., Yahya, M., Wahid, A., Yadav, P.,
  Breslin, J.G., Ali, M.I.: Tinyml benchmark: Executing fully connected neural
  networks on commodity microcontrollers. In: 2021 IEEE 7th World Forum on
  Internet of Things (WF-IoT). pp. 883--884. IEEE (2021)

\bibitem{sudharsan2020avoid}
Sudharsan, B., Sundaram, D., Breslin, J.G., Ali, M.I.: Avoid touching your
  face: A hand-to-face 3d motion dataset (covid-away) and trained models for
  smartwatches. In: 10th International Conference on the Internet of Things
  Companion. pp. 1--9 (2020)

\bibitem{sudharsan2021edge2guard}
Sudharsan, B., Sundaram, D., Patel, P., Breslin, J.G., Ali, M.I.: Edge2guard:
  Botnet attacks detecting offline models for resource-constrained iot devices.
  In: 2021 IEEE International Conference on Pervasive Computing and
  Communications Workshops and other Affiliated Events (PerCom Workshops). pp.
  680--685. IEEE (2021)

\bibitem{taylor_2019_over}
Taylor, C.: Over 50\% of irish people expected to own a smart speaker by 2023
  (04 2019),
  \url{https://www.irishtimes.com/business/technology/over-50-of-irish-people-expected-to-own-a-smart-speaker-by-2023-1.3869382}

\end{thebibliography}
\bibliographystyle{splncs03}

\end{document}